\documentstyle[12pt,aaspp4,flushrt]{article}
\begin{document}

\title{Constraining $\Omega$ with Cluster Evolution}

\author{Neta A. Bahcall, Xiaohui Fan, and Renyue Cen\\
Princeton Univeristy Observatory\\
Peyton Hall\\
Princeton, NJ 08544-1001\\
electronic mail : neta,fan,cen@astro.princeton.edu}

\section*{Abstract}
We show that the  evolution of the number density of rich clusters of galaxies
breaks the degeneracy between $\Omega$ (the mass density
ratio of the universe) and
$\sigma_{8}$ (the normalization of the power spectrum), $\sigma_{8} \:
\Omega^{0.5} \simeq 0.5$, that
follows from the observed present-day abundance of rich clusters.  The
evolution of high-mass
(Coma-like) clusters is strong in $\Omega = 1$, low-$\sigma_{8}$ models (such as
the standard biased CDM model with $\sigma_{8} \simeq 0.5$), where the
number density of clusters
decreases by a factor of $\sim 10^{3}$ from $z = 0$ to $z \simeq 0.5$; the
same clusters show only mild evolution in low-$\Omega$, high-$\sigma_{8}$ models,
where the decrease is
a factor of $\sim 10$.
This diagnostic provides a most powerful constraint on $\Omega$.
Using observations of clusters
to $z \simeq 0.5-1$, we find 
only mild evolution in the observed cluster abundance.
We find
$\Omega = 0.3 \pm 0.1$ and $\sigma_{8} = 0.85 \pm 0.15$
(for $\Lambda = 0$ models;
for $\Omega + \Lambda = 1$ models, $\Omega = 0.34 \pm 0.13$). 
These results imply, if confirmed by future surveys, that we live in a low-density,
low-bias universe.

{\em subject headings :} galaxies : clusters -- galaxies : evolution --
galaxies : formation -- cosmology : theory -- cosmology : observation --
dark matter 

\section{Introduction}
The observed present-day
abundance of rich clusters of galaxies 
places  one of the strongest constraints on cosmology 
(Bahcall and Cen 1992, White {\em et al.} 1993, Eke {\em et al.} 1996,
Viana and Liddle 1996, Pen 1996): $\sigma_{8} \: \Omega^{0.5}
\simeq 0.5 \pm 0.05$,
where $\sigma_{8}$ is the normalization of the power spectrum on
8 h$^{-1}$ Mpc scale (reflecting
the {\em rms} mass fluctuations on this scale),
and $\Omega$ is the present value of the cosmological density parameter. 
This constraint is degenerate in $\Omega - \sigma_{8}$; 
models with $\Omega = 1$ and $\sigma_{8}
\simeq 0.5$ are indistinguishable from models with $\Omega \simeq 0.25$ and
$\sigma_{8} \simeq 1$. (A
$\sigma_{8} \simeq 1$ universe implies no bias in the distribution of mass
versus light, since
$\sigma_{8}$(gal) $\simeq 1$ is observed for galaxies; a $\sigma_{8}
\simeq 0.5$ universe, on the
other hand, is highly biased, with mass distributed more
widely than light).

In the present paper, we show that a study of the evolution of the number
density of rich, massive
clusters as a function of redshift will  break the degeneracy between
$\Omega$ and $\sigma_{8}$ and
determine each parameter independently.  
The growth of high mass clusters depends strongly on the cosmology ---
mainly $\Omega$ and
$\sigma_{8}$ (e.g., Press and Schechter 1974, Peebles 1993, Cen and
Ostriker 1994a, Jing and Fang 1994, Eke {\em et al.}
1996, Viana and Liddle 1996).  In low-density models, density
fluctuations evolve and freeze
out at early times, thus producing only little evolution at recent times
($z \lesssim 1$). In an
$\Omega = 1$ universe, the fluctuations start growing only recently
thereby producing strong
evolution in recent times:  a large increase in the number density of
massive clusters is expected from $z \sim
1$ to $z = 0$. The evolution is so strong that finding even $\sim 1 - 2$
Coma-like mass clusters at $z
\simeq 0.5$ over $\sim 10^{3}$ deg$^{2}$ of sky would rule out an $\Omega = 1, 
\sigma_{8} \simeq 0.5$
model, where only $\sim 10^{-2}$ such clusters would be expected (\S 3).

We investigate in this paper the evolution of the mass function (MF) of clusters
(the number density of
clusters above a given mass) for various cosmologies using
large-scale N-body simulations,
and compare the results with cluster observations to
$z \simeq 0.5-1$.

\section{Model Simulations}
We investigate the evolution of the cluster MF in five cosmological models using
large-scale numerical simulations.
The models include: Standard Cold Dark Matter (SCDM; $\Omega =
1$), normalized to a
present-day mass fluctuation on 8h$^{-1}$ Mpc scale of $\sigma_{8} =
1.05$ (consistent with the
COBE microwave background fluctuations on large scales); a biased SCDM
($\Omega = 1$), with low
normalization ($\sigma_{8} = 0.53$), which fits the present day cluster abundance 
(but is
inconsistent with the COBE normalization);  a low-density, $\Lambda$ dominated
CDM model (LCDM), and an
open CDM model (OCDM); and a mixed, hot and cold dark matter model
 (MDM, $\Omega$=1, $\Omega_{\nu}$=0.3;
from Cen
and Ostriker 1994b, normalized to $\sigma_{8}$=0.6, 
consistent with the present day cluster adundance).
 Table 1 summarizes the model parameters.
All models, except for the
COBE-normalized SCDM ($\sigma_{8} \simeq 1$), are 
consistent with the present day
cluster abundance (Bahcall and Cen 1992, White {\em et
al.} 1993, Cen and Ostriker 1994b, Eke {\em et
al.} 1996, Pen 1996). The SCDM $\sigma_{8} \simeq 1$ model over-produces
the number of massive
clusters by an order of magnitude. All models except
for the biased SCDM
($\sigma_{8} \sim 0.5$) are also consistent with the COBE normalization (Bunn \& White 1996).

A large-scale particle-mesh code with box size 400 $\rm h^{-1}$ Mpc was
used to simulate the evolution
of the dark matter in the models.  A large simulation box is needed in
order to produce a significant
number of the rich but rare clusters ($\rm \lesssim 10^{-5}h^{3}$ clusters
$\rm Mpc^{-3}$). The simulation
box contains $720^{3}$ cells and $240^{3} = 10^{7.1}$ dark matter
particles, with a particle mass of $1.3 \times 10^{12} \Omega \rm h^{-1} \rm M_{\odot}$. 
In each simulation, clusters are selected as the maxima of the mass distribution  within
spheres of comoving radius of $1.5\rm h^{-1}$ Mpc.
The mass of each cluster is determined within two relevant
radii: a co-moving radius
of $\rm R_{\rm com} = 1.5\rm h^{-1}$ Mpc, and a physical radius of $\rm R_{\rm
{phy}} = 1.0\rm h^{-1}$ Mpc. 
We use these radii in order to allow a
proper comparison with
observations, which generally employ $\rm R_{\rm{com}}$ or $\rm R_{\rm{phy}}$ as
their observable parameter.
A virial cluster radius, which is commonly used in theoretical analyses
(such as the Press-Schechter
approximation), generally cannot be accurately determined from observations.
To study the evolution of the cluster mass function, cluster masses are
determined at several
redshifts: $z$ = 0, 0.5, 1 and 2. The cluster MF, n($>$M), which
represents the number density
of clusters above a given mass threshold is then determined at each
redshift. The evolution of the
cluster mass function is derived for each model and compared with observations.
\vspace{-0.7cm}
\section{Evolution of the Cluster Mass Function}
The evolution of the cluster MF is presented in Fig. 1 for
two representative models : biased $\Omega$=1 SCDM and low-density
OCDM, which are ``degenerate'' at z = 0.
A negative evolution of the
cluster MF is seen in all
models --- i.e., the abundance of clusters decreases at earlier epochs for clusters 
 of a given mass, 
since massive clusters grow with time (e.g., Press and Schechter 1974, Peebles
{\em et al.} 1989, Peebles
1993, Cen and Ostriker 1994a, Luppino and Gioia 1995, Eke {\em et al.} 1996).
The {\em rate} of the evolutionary growth, however, is strongly model dependent.
For example, the number
density of $\rm M(\leq 1.5) \geq 3 \times 10^{14}\rm h^{-1} \rm M_{\odot}$  clusters
drops by a factor of $\sim 40$ from $z = 0$ to $z =
0.5$ in the biased SCDM
cosmology, while the drop is only a factor of 4 (instead of 40) in the
low-density models (OCDM, LCDM).
The difference becomes even larger for more
massive clusters (see below). 

The evolution of the cluster density as a function of redshift is presented
in Fig. 2 for massive clusters
with mass $\rm M(\leq R_{\rm{com}} = 1.5h^{-1} \rm Mpc)\geq 5.5 \times
10^{14}\rm h^{-1} \rm M_{\odot}$
(corresponding to richness class $\gtrsim 2.5$; Bahcall and Cen 1993),
and in Fig. 3 for less massive clusters $\rm M(\leq R_{\rm{phy}} = 1.0h^{-1} \rm Mpc)\geq 1.5 \times
10^{14}\rm h^{-1} \rm M_{\odot}$ (corresponding to richness class $\gtrsim 0$,
see also \S 4).
At $z \simeq 0$, all models
except $\sigma_{8} \simeq 1$ SCDM yield a comparable abundance of clusters,
consistent with
observations.
 The $\sigma_{8} \sim 1$ SCDM model produces an
order-of-magnitude more clusters than observed. At high redshifts,
the abundance of clusters
decreases sharply for the low $\sigma_{8}$ models 
but the decrease is slow for higher $\sigma_{8}$.
The evolution rate is insensitive to the value of the Hubble constant,
or the exact shape of the power spectrum,
and is most sensitive
to the normalization $\sigma_{8}$ (for same mass clusters).
The dependence on $\Omega$
itself is in fact only secondary. The strong exponential dependence on $\sigma_{8}$
results from the fact that for a given {\em mass} cluster, a
lower $\sigma_{8}$ implies
the clusters are rarer peaks in the density distribution 
, therefore evolving considerably faster than in high  $\sigma_{8}$ models
(see Fan {\em et al.} 1997).
An observational determination of the cluster evolution rate therefore enables
us to directly constrain $\sigma_{8}$.

\section{Comparison with Observations}
Systematic observations of clusters of galaxies at high resdhifts are only
now beginning, with the use of 
complete redshift surveys (determining cluster mass from 
velocity dispersion), X-ray
observations (temperatures of  clusters), and weak gravitational
lensing. New
complete surveys of optical and x-ray clusters at low to high redshifts ($z
\gtrsim 0.5)$
will become available over the next several years. Here we
present results from two
independent current optical cluster surveys in the redshift range $z \simeq
0$ to  $\sim 1$. While the
current samples are still small and the uncertainties large, the
sensitive cluster evolution already
allows us to place strong constraints on the cosmology.

The CNOC optical cluster redshift survey (Carlberg
{\em at al.} 1996) 
represents a small but complete redshift survey of high mass
clusters in the redshift
range $z = 0.18 - 0.55$, with an EMSS  extension at $z = 0.55 - 0.85$
(Henry {\em et al.} 1992,
Luppino and Gioia 1995, Carlberg {\em et al.} 1997). Redshifts for
typically $\sim 30$ to $>
100$ galaxies per cluster are used to accurately determine the velocity
dispersion and mass of each
cluster (Carlberg {\em et al.} 1996). The cluster mass threshold and
cluster densities in the survey,
properly corrected for completeness effects, are discussed by Carlberg {\em et
al.} (1997). The mass
threshold used is based on a velocity dispersion threshold
of $\sigma_{r} \geq 800 \rm km
s^{-1}$, which corresponds to a mass (within $R_{\rm{com}} =
1.5\rm h^{-1}$ Mpc) of $\rm M(\leq
1.5) \geq 5.5 \times 10^{14}h^{-1} M_{\odot}$ 
(as determined from the CNOC data as well as compared with Coma).
 The cluster densities above
this threshold are $3.5 \times
10^{-7} \rm Mpc^{-3}$ at $z = 0.18-0.35$, and $9.3 \times
10^{-8} \rm Mpc^{-3}$ at $z = 0.35-0.55$ (for $\Omega = 1$).
The high redshift extension at $z = 0.55-0.85$ (Luppino and Gioia 1995,
Carlberg {\em et al.} 1997)
does not have complete velocity measurements 
but contains the richest and most luminous X-ray clusters
(some with observed velocity dispersions $\sigma_{r} \gtrsim 1200 \rm km \rm)$.
We conservatively assume that these clusters have the same mass threshold 
as the CNOC clusters (also selected from EMSS).
The mass threshold is likely to be higher; if so, this will 
raise the best-fit $\sigma_{8}$ value. 
This uncertainty is included in our estimates.
The observed abundance of nearby
clusters $(z \simeq 0-0.1)$ is taken from Bahcall and Cen (1993),
Mazure {\em et al.} (1996, the ESO survey) and 
Henry and Arnaud (1992;
based on X-ray cluster selection), 
all converted to the common mass threshold of $5.5 \times 10^{14}\rm h^{-1} \rm M_{\odot}$  (using
the observed MF, Bahcall and Cen 1993). This
common mass threshold 
is slightly lower than a Coma-type cluster 
($\sim 6.5 \times 10^{14} \rm h^{-1} \rm M_{\odot}$; Hughes 1989, Bahcall \& Cen
1993).
The sensitivity of the results to the exact mass threshold 
is tested by varying the assumed threshold from
5 to 6.5 $\times 10^{14} \rm h^{-1} \rm M_{\odot}$;
the results (\S 5) include these uncertainties.

The results are presented in Fig. 2, together with the model expectations
for this mass threshold clusters.
(The model $\sigma_{r}$ and mass thresholds are corrected for the resolution effect
 of the simulation by comparing with high resolution simulations; the effect
is small: $\lesssim 5\%$). 
Only a mild negative evolution is observed. This mild evolution is
in excellent agreement
with the low-density high-normalization models (OCDM, LCDM); it is 
inconsistent by a factor of
10 -- 100 with the very strong evolution expected in the biased
($\sigma_{8} \simeq 0.5)$ SCDM and MDM
models. The expected cluster density decreases by a factor of $\sim 10$ from
$z = 0$ to $z = 0.5$ in
the low-density models, while the decrease becomes enormous $(\sim 10^{3})$
for $\sigma_{8} \simeq
0.5$ SCDM, and $\sim 10^{2}$ for MDM. The data show a decrease by a factor of $\sim 10$ to $z
\simeq 0.5$. This
comparison differentiates the $\Omega = 1$, $\sigma_{8} \simeq 0.5$
models from the $\Omega
\simeq 0.3-0.4 ,\sigma_{8} \simeq 0.8$ models, which are
indistinguishable at $z \simeq 0$.
Only the low-density, higher normalization models
are acceptable at high
redshifts (see \S 5). The unbiased $\sigma_{8} \simeq 1$ SCDM model, which also yields
mild evolution (due to
its high $\sigma_{8}$), is inconsistent with the observed cluster
abundance at any redshift.

The second cluster sample we investigate is the Palomar Distant Cluster Survey
(PDCS; Postman {\em et al.}
1996). The PDCS is a complete automated survey of distant clusters to $z
\sim 1$ from deep imaging CCD
data over 5 deg$^{2}$. Clusters were selected from the imaging data using a
matched-filter algorithm,
which yields best-fit estimates of the cluster richness ($\propto$
luminosity) and redshift. While the
clusters do not have measured redshifts and velocity dispersions (i.e.,
masses), the estimated
luminosities are determined in a consistent manner from $z
\simeq 0.2$ to $\sim 1$, enabling
us to investigate the evolutionary trend of the cluster densities.
Measurements of the cluster
redshifts and velocity dispersions will eventually provide more accurate
results.
We select all clusters with luminosities $L_{cl}
 \geq 50L^{*}$
(selected in the $I$-band, with richness threshold $\Lambda_{cl} \geq 50$
where $L_{cl} =
\Lambda_{cl}L^{*}$ within a physical radius of $1\rm h^{-1}$ Mpc, the radius
used by the PDCS
selection). This corresponds to a conservative mass threshold of
$M(R_{\rm{phy}} = 1 \rm h^{-1} \rm Mpc)
\gtrsim 1.5 \times 10^{14}\rm h^{-1} \rm Mpc$ for an average cluster $M/L \sim
300\rm h$ (Bahcall {\em et
al.} 1995, Carlberg {\em et al.} 1996). (The evolution results are not
sensitive to the exact 
mass threshold at these low mass values; see below).

Figure 3 presents  the evolution of the PDCS cluster density to $z \simeq 1$.
The most distant point, at $z \simeq 0.9$, includes an incompleteness
correction using the PDCS
calibrated selection function at that redshift, and is thus less accurate.
(The error bar includes
both statistical uncertainties and a conservatively estimated uncertainty due
to the selection
correction).
Figure 3 compares the observed evolution with the model expectations
for the same physical
radius and mass threshold clusters. The data, again, show only a
minimal evolution of the cluster
density, in excellent agreement  with the low-density, high-$\sigma_{8}$
models. The data are 
inconsistent with the biased SCDM and MDM models, which predict $\sim 10$ times lower
cluster density than
observed at $z \sim 1$. (If
the actual mass threshold
of the clusters is larger than estimated above, then the model evolution 
will be even stronger and the evolutionary difference among
models 
somewhat larger;
in that respect the assumed mass threshold is a
conservative choice).

The results are similar for the PDCS and the CNOC samples.
While the CNOC clusters
represent considerably higher mass clusters, 
which are most sensitive to the cosmology,
the PDCS clusters reach to higher redshifts of $z \sim 1$.
The fact that both independent samples, with different mass threshold clusters 
and different selection algorthiums,
yield similar results provides further support to these conclusions.

\section{Constraining $\Omega$}
%

 A comparsion of the observed cluster evolution with the
models shows that the data are 
consistent with the low-density models (OCDM and LCDM), and are inconsistent
with the $\Omega = 1$ models (SCDM, biased SCDM and MDM).
The relatively mild evolution observed in both the CNOC and PDCS samples 
is consistent with OCDM 
at a significance level of $\sim 60\%$  
(based on a $\chi^{2}$ test), and
with LCDM at $\sim 30\%$.
The $\Omega = 1$ SCDM and MDM models are rejected at $> 99.9\%$.

 We use the data to directly determine the best-fit values of $\Omega$ and
$\sigma_{8}$ for the CDM models. We use the method described by Fan {\em et al.} (1997),
correlating the evolution rate (n(z)/n(0)) with $\sigma_{8}$ to determine
$\sigma_{8}$ directly, since the primary dependence of the evolution rate
is on $\sigma_{8}$ (for same mass clusters). The evolution rate is 
 exponentially dependent on $\sigma_{8}^{2}$ -- increasing strongly as $\sigma_{8}$
decreases; it is nearly independent of other parameters, including $\Omega$
(see Fan {\em et al.}). We determine the best-fit relation for the evolution rate 
versus $\sigma_{8}$ from the model simulations and compare the expected
relation with the observed evolution rate. We find $\sigma_{8} =
0.85 \pm 0.15$. The observed mild evolution rate of rich clusters thus implies
a nearly unbiased universe; a strongly biased universe ($\sigma_{8} < 0.7$) is unlikely 
since it produces considerably stronger evolution than observed. 
The results are consistent with those of Carlberg {\em et al.} (1997) of
$\sigma_{8} = 0.75 \pm 0.1$.
Combined with the $\Omega - \sigma_{8}$ relation for present-day cluster
adundance for CDM models (from Eke {\em et al.} 1996), we find $\Omega = 0.3 \pm 0.1$ 
(for $\Lambda$ = 0), and $\Omega = 0.34 \pm 0.13$ ( for $\Lambda = 1 - \Omega$).  
The results are presented in Fig. 4. 
The figure illustrates the powerful diagnostic of cluster evolution in 
determing $\Omega$ and $\sigma_{8}$; it places the strongest constraints
yet on these parameters.
The independent constraint placed by cluster dynamics, $\Omega \simeq (0.2 \pm
0.07)\sigma_{8}^{-1}$ (assuming linear bias; Bahcall {\em et al.} 1995, Carlberg {\em et al.}
1996, 1997) is also shown in
the figure; it is  consistent with the above results, yielding 
$\Omega = 0.24 \pm 0.1$ for $\sigma_{8} = 0.85 \pm 0.15$. 
This $\Omega$ range provides the overlap of the constraints placed
by the cluster abundance evolution and cluster dynamics observations.
These results suggest that we live in a low-density, low-bias universe.
Recent observations  suggesting a minimal negative evolution of the X-ray cluster luminosity function 
(Henry {\em et al.} 1992, Castander {\em et al}, 1994, Collins {\em et al.} 1997,
Nichol {\em et al.} 1997, Mushotzsky {\em et al.} 1997)
are also consistent with the above findings.

 We thank J.P.Ostriker, J.P.E.Peebles, D.N.Spergel and M.A.Strauss for helpful discussions.
This work is supported by NSF grants AST93-15368, ASC93-18185 and NASA grant NAG5-2759.
XF thanks the support from an Advisory Council Scholarship. 

\newpage
%

\newpage
Table 1. Model Parameters

\begin{center}
\begin{tabular}{cccccc}\hline \hline
     & $\Omega$ & $\Lambda$ & h & $\sigma_{8} $ \\ \hline
SCDM & 1.0      & 0.0   &0.5    &1.05  \\
SCDM & 1.0      & 0.0   & 0.5   & 0.53  \\
MDM  & 1.0      & 0.0   & 0.5   & 0.60 \\
OCDM & 0.35     & 0.0   & 0.7   & 0.80 \\
LCDM & 0.4      & 0.6   & 0.65  & 0.79 \\ \hline \hline

\end{tabular}
\end{center}
\newpage
\begin{figure}
\vspace{-6cm}

\epsfysize=600pt \epsfbox{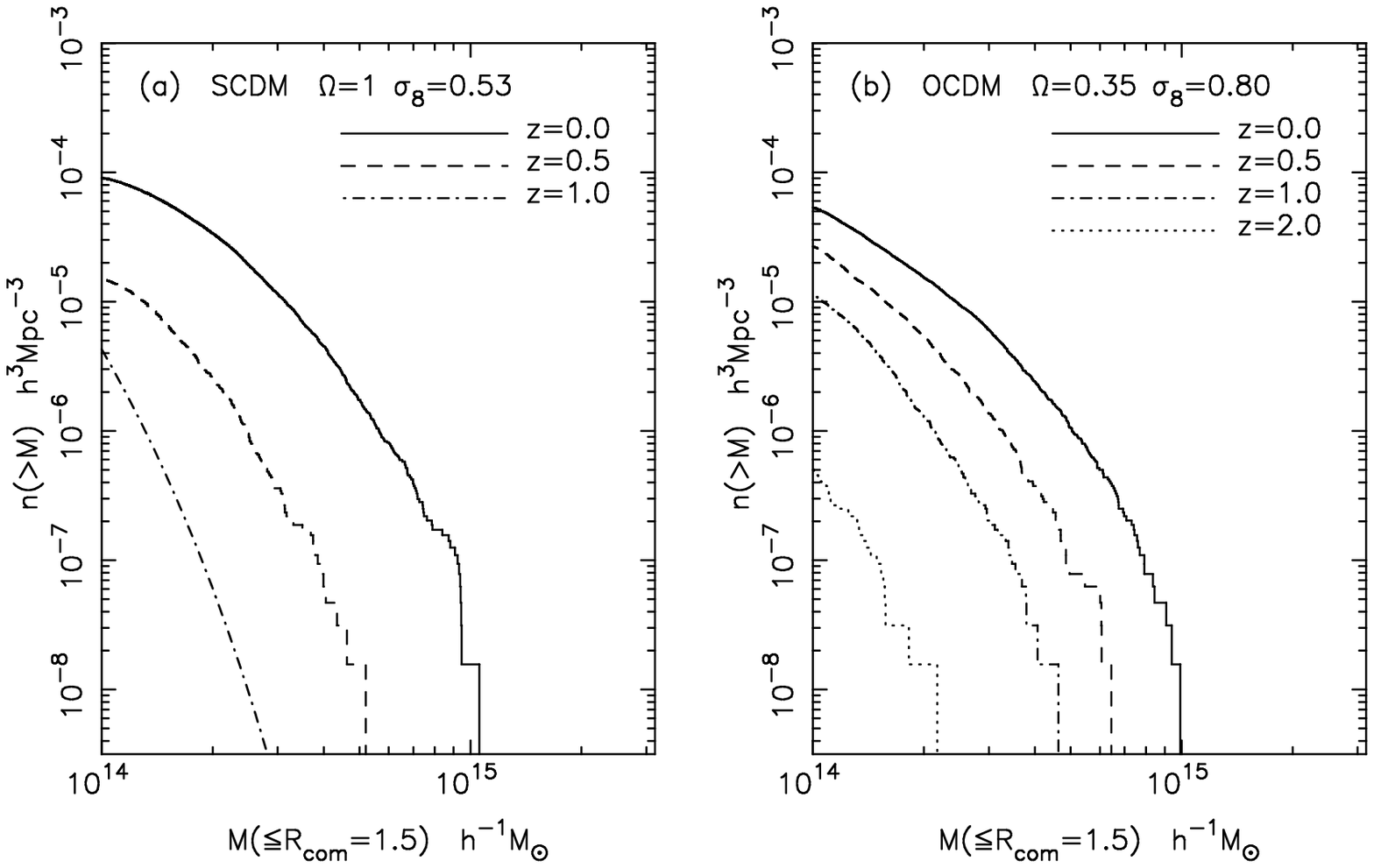}

\vspace{3cm}
Figure 1. The evolution of the cluster mass function with redshift, for  
cluster masses within a co-moving radius $R_{\rm com}$ = 1.5$\rm h^{-1}$ Mpc.
(For z=1 in SCDM $\sigma_{8}$ = 0.53, Press-Schechter appoximation is used.)
\end{figure}
\begin{figure}
\vspace{-6cm}

\epsfysize=600pt \epsfbox{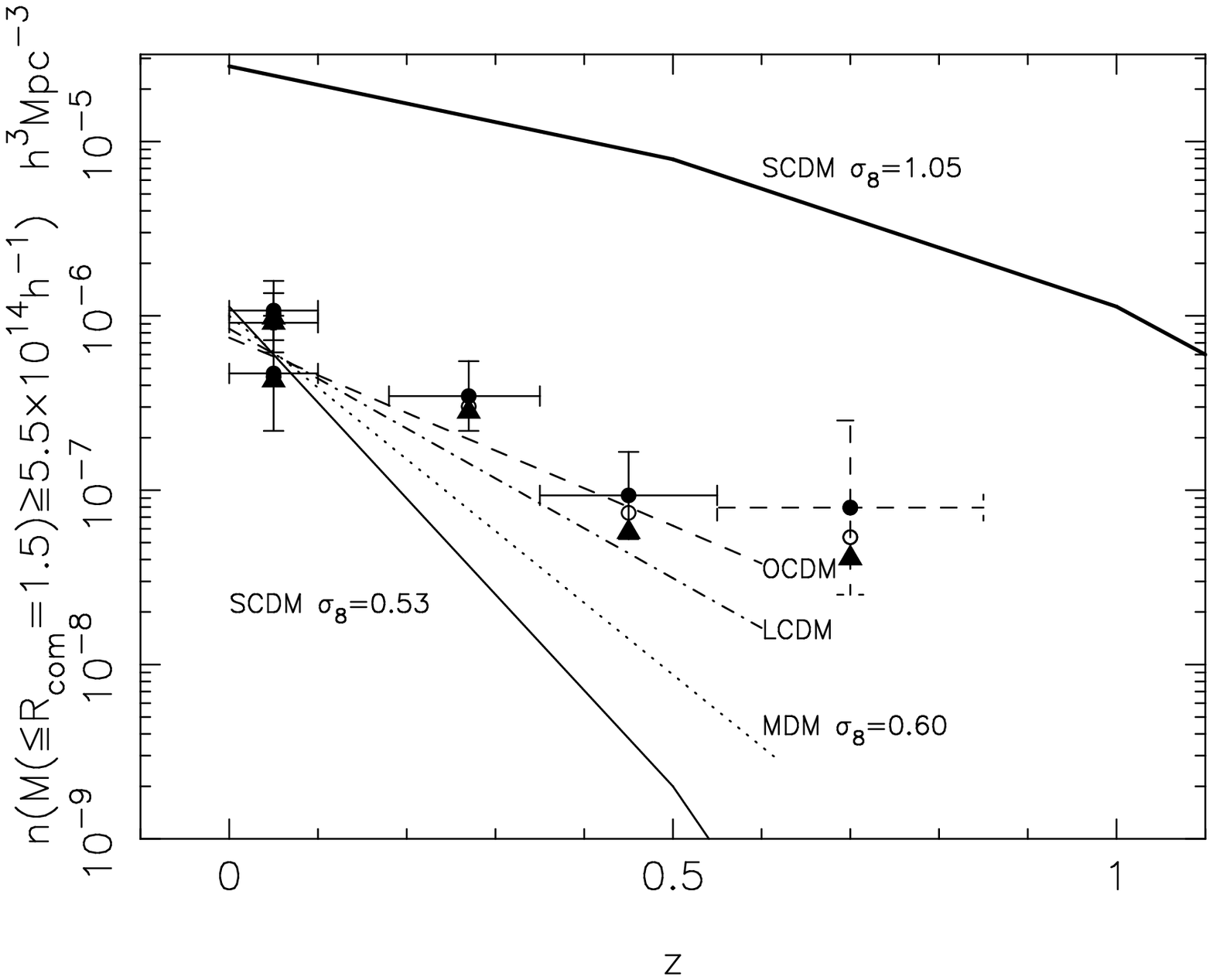}

\vspace{3cm}
Figure 2. Observed vs. model  cluster abundance
as a function of redshift for clusters with  mass
$\rm M(\leq \rm R_{\rm com}= 1.5 \rm h^{-1} \rm Mpc) \geq 5.5 \times 10^{14}\rm h^{-1} \rm M_{\odot}$.
The observed abundance  at z $\sim$ 0 are from 
 Bahcall and Cen (1992),
Mazure {\em et al.} (1996) and Henry and Arnaud (1992).
The data at z $\sim$ 0.27 and 0.45 are from the CNOC survey (Carlberg {\em et al.}
 1997),
and at z $\sim$ 0.7 from Luppino and Gioia (1995). 
The different symbols represent the observed number densities for 
$\Omega$=1 (filled circles), $\Omega$=0.35, $\Lambda$=0 (open circles), and $\Omega$=0.4, $\Lambda$=0.6 (triangles).
\end{figure}
\begin{figure}
\vspace{-6cm}

\epsfysize=600pt \epsfbox{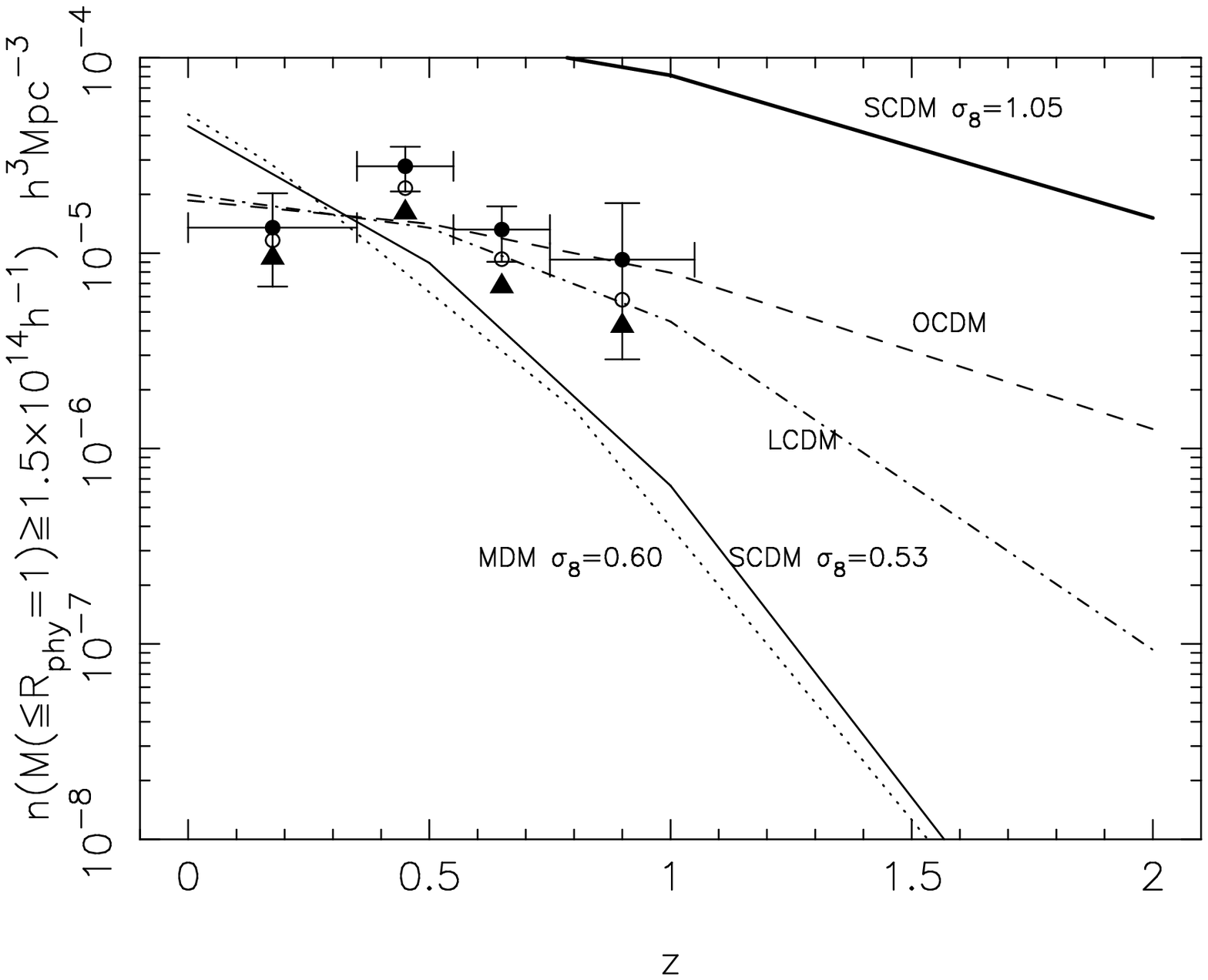}

\vspace{3cm}
Figure 3. Observed vs. model  cluster abundance
as a function of redshift for clusters with  mass
$\rm M(\leq \rm R_{\rm phy}= 1.0 \rm h^{-1} \rm Mpc) \geq 1.5 \times 10^{14}\rm h^{-1} \rm M_{\odot}$.
The data  are from the PDCS survey (Postman {\em et al.}, 1996). The different symbols
for the data are  the same as in Fig. 3.
\end{figure}
\begin{figure}
\vspace{-6cm}

\epsfysize=600pt \epsfbox{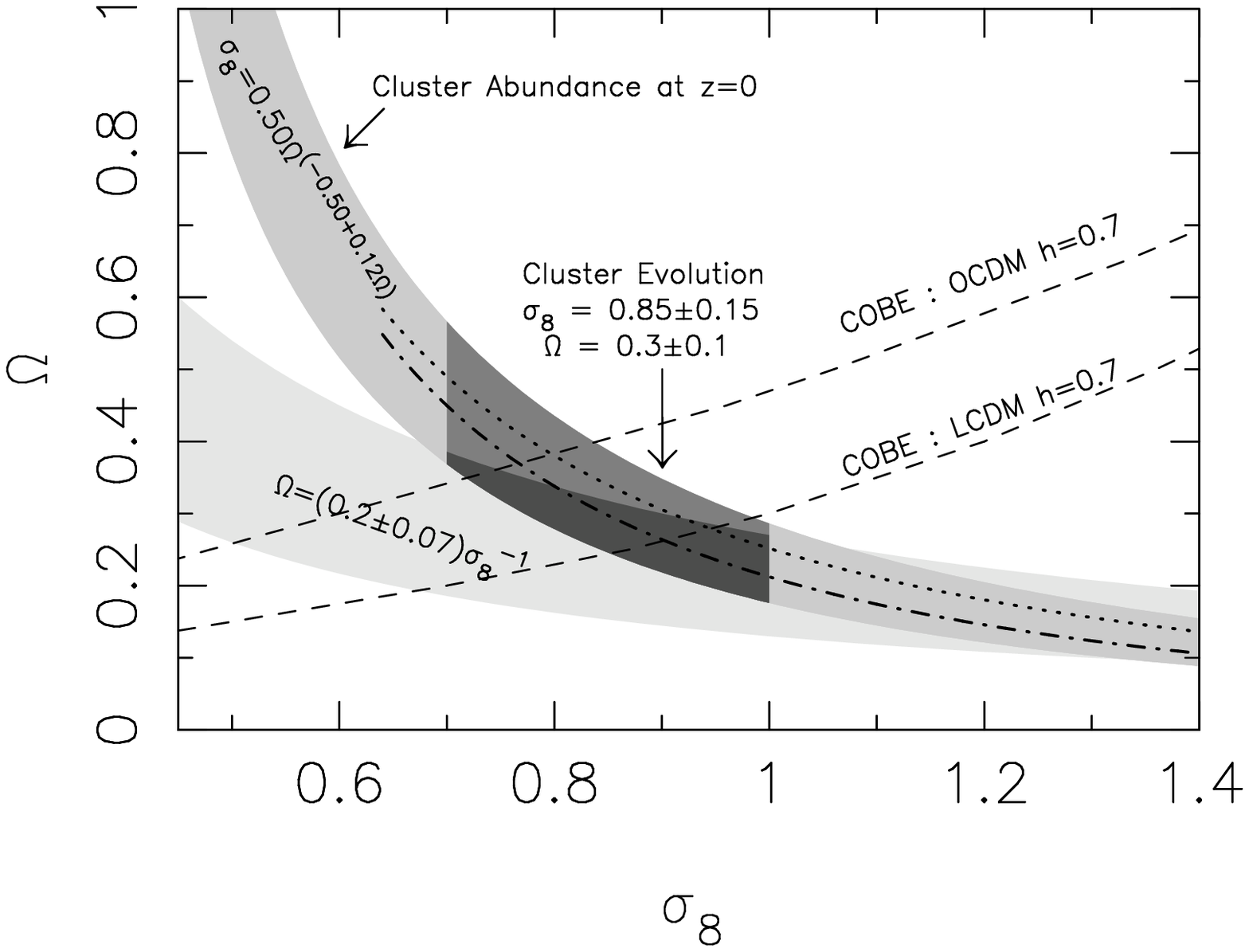}

\vspace{1cm}
Figure 4. Observational constraints on $\Omega$ and $\sigma_{8}$.
The band $\sigma_{8} = 0.50 \Omega^{-0.50+0.12\Omega}$ represents  the
range due to the present day cluster abundance (for the average of
open and $\Lambda$ models, Eke {\em et al.} 1996; the inserted dash-dotted and dotted
 lines
are the best fits for open and $\Lambda$ models, respectively).
The darker band of $\sigma_{8} = 0.85 \pm 0.15$, $\Omega = 0.3 \pm 0.1$  is the const
raint
placed in this paper by cluster evolution (\S 5).
The $\Omega = (0.2\pm0.07) \sigma_{8}^{-1}$ band represents cluster
dynamics constraint.
The dashed lines are the COBE four year data
(Bunn and White 1996).
A low-density low-bias
universe with $\Omega = 0.3 \pm 0.1$ and $\sigma_{8} = 0.85 \pm 0.15 $ best fits
all the data (darkest region), with cluster evolution providing the tightest constraint.
\end{figure}
\end{document}